\documentclass[prl,twocolumn,a4paper,superscriptaddress]{revtex4}
\usepackage[english]{babel}
\usepackage{amsmath}
\usepackage{amsfonts}
\usepackage{amssymb}
\usepackage{array}
\usepackage{dcolumn}
\usepackage{longtable}
\usepackage{hyperref}
\usepackage{comment}
\usepackage{float}
\usepackage{bbm}
\usepackage{bm}
\usepackage{graphicx}
\usepackage{natbib}

\begin{document}
\title{Geometrical frustration yields fiber formation in self-assembly}
\date{\today}
\author{Martin Lenz}
\email{martin.lenz@u-psud.fr}
\affiliation{LPTMS, CNRS, Univ. Paris-Sud, Universit\'e Paris-Saclay, 91405 Orsay, France}
\author{Thomas A. Witten}
\affiliation{Department of Physics and James Franck Institute, University of Chicago, Chicago, Illinois 60637, United States.}

\maketitle

{\bf
Controlling the self-assembly of supramolecular structures is vital for living cells, and a central challenge for engineering at the nano- and microscales~\cite{Glotzer:2007,McManus:2016}. 
Nevertheless, even particles without optimized shapes can robustly form well-defined morphologies. This is the case in numerous medical conditions where normally soluble proteins aggregate into fibers~\cite{Eaton:1990,Knowles:2014}. Beyond the diversity of molecular mechanisms involved~\cite{Nelson:2006a,Eichner:2011}, we propose that fibers generically arise from the aggregation of irregular particles with short-range interactions. Using a minimal model of ill-fitting, sticky particles, we demonstrate robust fiber formation for a variety of particle shapes and aggregation conditions. Geometrical frustration plays a crucial role in this process, and accounts for the range of parameters in which fibers form as well as for their metastable character.
}

Identical cubes can pack into dense space-filling aggregates, but most shapes do not. As a result, the aggregates formed by these shapes tend to be frustrated, giving rise to arrested, glassy states~\cite{Foffi:2002,Cardinaux:2007}.
In protein aggregates, this frustration can arise from, \emph{e.g.}, deformed or partially denatured protein domains, the juxtaposition of residues with unfavorable interactions or sterically hindered hydrogen bonding.
Any compact packing of these objects thus involves tradeoffs between geometrical constraints, which  hinder the formation of compact aggregates, and the particles' overall attractive interactions.
As a result, the global morphology of the aggregate is controlled by the competition between these two effects.

\begin{figure}[b]
\centering
\includegraphics[width=\columnwidth]{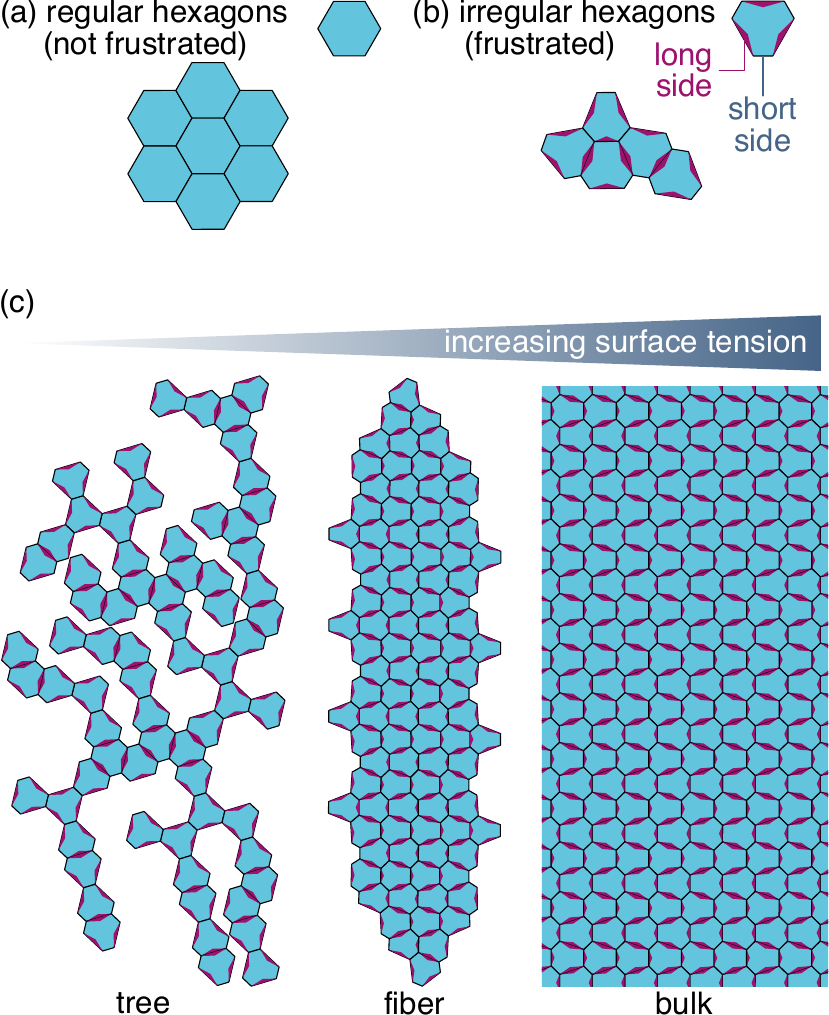}
\caption{\label{fig:model}
\textbf{Frustration and adhesion compete to determine the morphology of aggregates of mismatched particles.}
(a)~Polygonal particles with well-matched shapes (here regular hexagons) readily aggregate into space-filling two-dimensional aggregates.
(b)~Generic particles (\emph{e.g.}, irregular hexagons with long and short sides) must be distorted to form a compact aggregate, resulting in geometrical frustration.
(c)~At low surface tensions, frustration precludes the formation of compact polygon packings, yielding tree-like aggregates. Conversely, a large surface tension results in a bulk. Fibers constitute a compromise between these two extremes.}
\end{figure}

To explore this competition in its simplest form, we consider two-dimensional, deformable polygons driven to aggregate by zero-range attractive interactions [Fig.~\ref{fig:model}(a-b)]. We parametrize the magnitude of this attraction by a surface tension whose value controls the morphology of the aggregate [Fig.~\ref{fig:model}(c)]. A low surface tension thus favors thin tree-like aggregates composed of undeformed particles with very little elastic frustration, reminiscent of so-called empty liquids~\cite{Bianchi:2006}. Conversely, a large surface tension leads to space-filling aggregates in which all particles are substantially deformed. In this paper, we demonstrate that fibers form at intermediate values of the tension, where the characteristic energies associated with particle attraction and deformation are comparable. We quantitatively account for these values based on the role of frustration, and show that fibers are very robust to changes in microscopic parameters, aggregation protocol and seeding conditions. Finally, we show that despite this robustness fibers do not constitute the ground state of our aggregates. Instead, they are kinetically trapped metastable states, consistent with their inherent frustration and with the well-documented irreversible character of protein fiber assembly \emph{in vivo}.

\begin{figure}[t]
\centering
\includegraphics[width=\columnwidth]{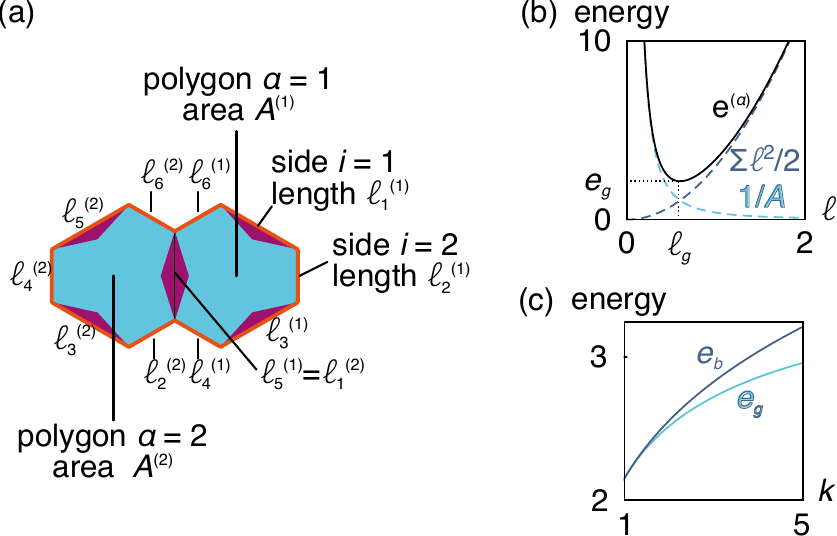}
\caption{\label{fig:parametrization}\textbf{Polygon model of aggregation.} Here we present hexagons with sides of alternating length characterized by an asymmetry parameter $k>1$.
(a)~Two-hexagon aggregate with two joined sides and $10$ unjoined sides outlined in \emph{orange}.
(b)~Magnitude of the total energy $e^{(\alpha)}$ of a $k=1$ regular hexagon of side $\ell$. We also plot separately each of its two terms as defined in the Methods. The equilibrium side length $\ell_g$ and ground-state deformation energy $e_g$ are shown.
(c)~Hexagons with alternating sides become increasingly frustrated as their asymmetry is increased ($k\neq 1$), as evidenced by the increasing gap between the energies of the ground state ($e_g$) and the bulk [$e_b$, corresponding to the bulk topology of Fig.~\ref{fig:model}(c)].
}
\end{figure}

We consider $n$-sided polygons, and use a deformation energy $e^{(\alpha)}$ for the $\alpha$th polygon that is a function of its area $A^{(\alpha)}$ and of the lengths $\lbrace \ell_i^{(\alpha)}\rbrace_{i=1\ldots n}$ of its sides as shown in Fig.~\ref{fig:parametrization}(a). In the following we use both regular and irregular polygons [as in Fig.~\ref{fig:model}(b)], the latter being characterized by an asymmetry parameter $k\geq 1$, where $k\rightarrow 1$ is the regular polygon limit while $k\rightarrow +\infty$ yields short sides with vanishing spontaneous length (see Methods). Minimizing the energy $e^{(\alpha)}$ with respect to the positions of the polygon's vertices yields a rigid elastic ground state of energy $e_g$ [Fig.~\ref{fig:parametrization}(b)]. Aggregates are formed by connecting multiple polygons through the joining of one or several of their sides. Two joined sides are treated as a single object, implying that they share the same two end-vertices  [Fig.~\ref{fig:parametrization}(a)]. Side joining is favored by the adhesion energy between particles, modeled by an energy penalty $\sigma>0$ for each unjoined side regardless of its actual length [$\sigma$ thus parametrizes the surface tension introduced in Fig.~\ref{fig:model}(c)]. However, it also involves a distortion of the mismatched polygons, and thus increases their deformation energy above $e_g$.

A tree such as the one of Fig.~\ref{fig:model}(c) is always in its elastic ground state. Its average energy per particle is thus entirely due to surface tension, and reads $e\sim(n-2)\sigma$ in the thermodynamic limit (see Methods). A bulk, on the other hand, has a negligible surface energy in the thermodynamic limit but a finite deformation energy $e\sim e_b-e_g$, where $e_b$ denotes the minimal deformation energy for a polygon constrained by the bulk topology [Fig.~\ref{fig:parametrization}(c)]. Rescaling the elastic energy and tension through $\tilde{e}^{(\alpha)}=[e^{(\alpha)}-e_g]/(e_b-e_g)$ and $\tilde{\sigma}=(n-2)\sigma/(e_b-e_g)$, we obtain $\tilde{e}=\tilde{\sigma}$ for a tree and $\tilde{e}=1$ for a bulk. This rescaled form for the energy makes it clear that adhesion overcomes frustrations and trees become less stable than bulks at high tensions, with a transition at $\tilde{\sigma}=1$. As a result, if fibers indeed form as a result of the competition between these two effects, we expect them to appear for a dimensionless tension of order one.

\begin{figure*}[t]
\centering
\includegraphics[width=.99\textwidth]{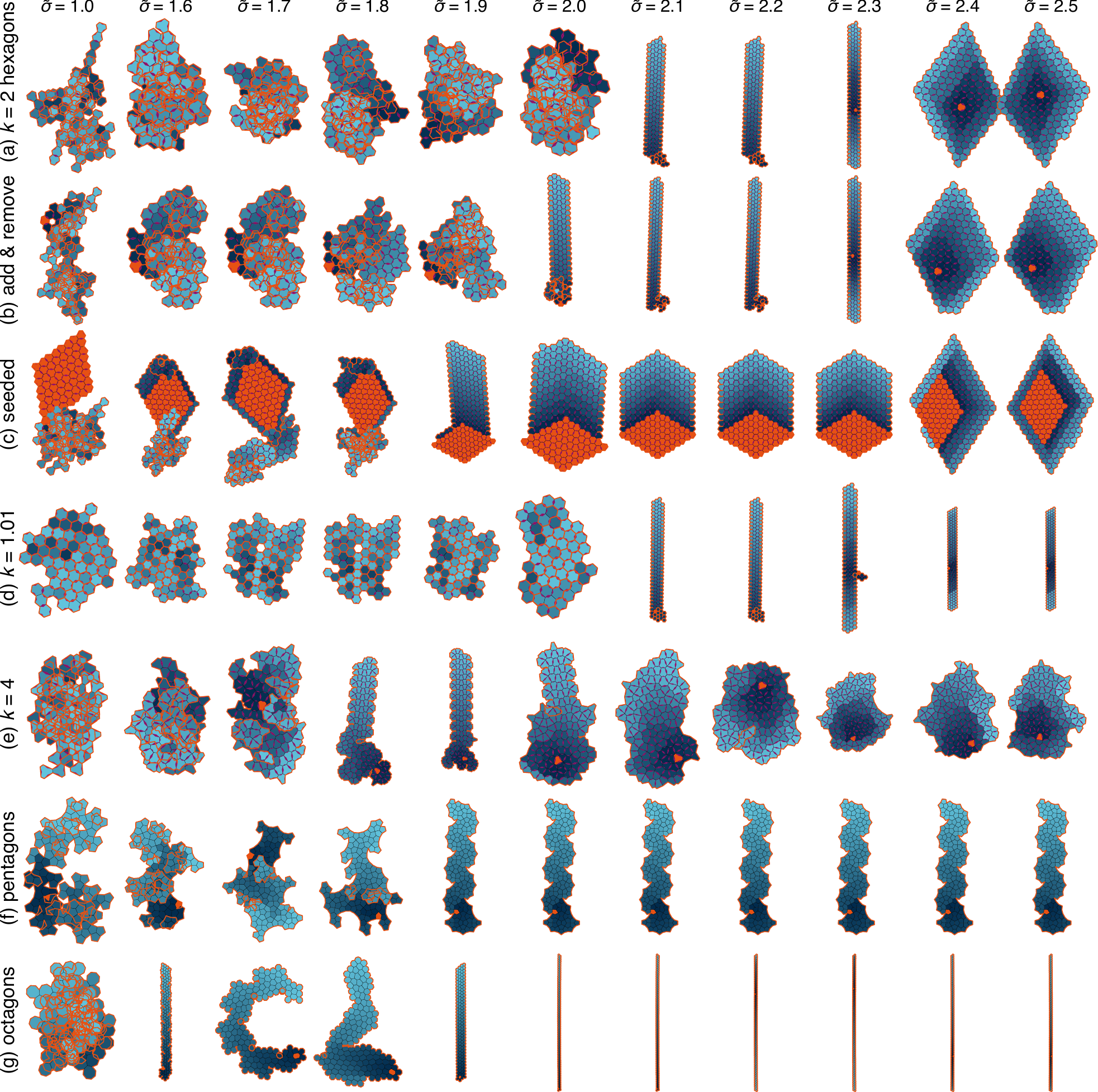}
\caption{\label{fig:fibers}\textbf{Aggregates resulting from our growth algorithm.} For all conditions probed (shown on figure), periodic fibers form at intermediate rescaled surface tension $\tilde{\sigma}$. \emph{Orange polygons} represent the initial topology used to initiate the growth algorithm. Older polygons are colored in \emph{dark blue}, while \emph{light blue} denotes the most recent additions. The \emph{orange line} outlines unjoined sides. Trees form at low tension, but superficially appear much more compact than the illustration of Fig.~\ref{fig:model}(c): this is because their many branches overlap each other, as can be seen from the convoluted, re-entrant orange aggregate boundary. Such self-overlaps are not penalized in our algorithm. Bulks form at high tensions.
}
\end{figure*}

\begin{figure*}[t]
\centering
\includegraphics[width=\textwidth]{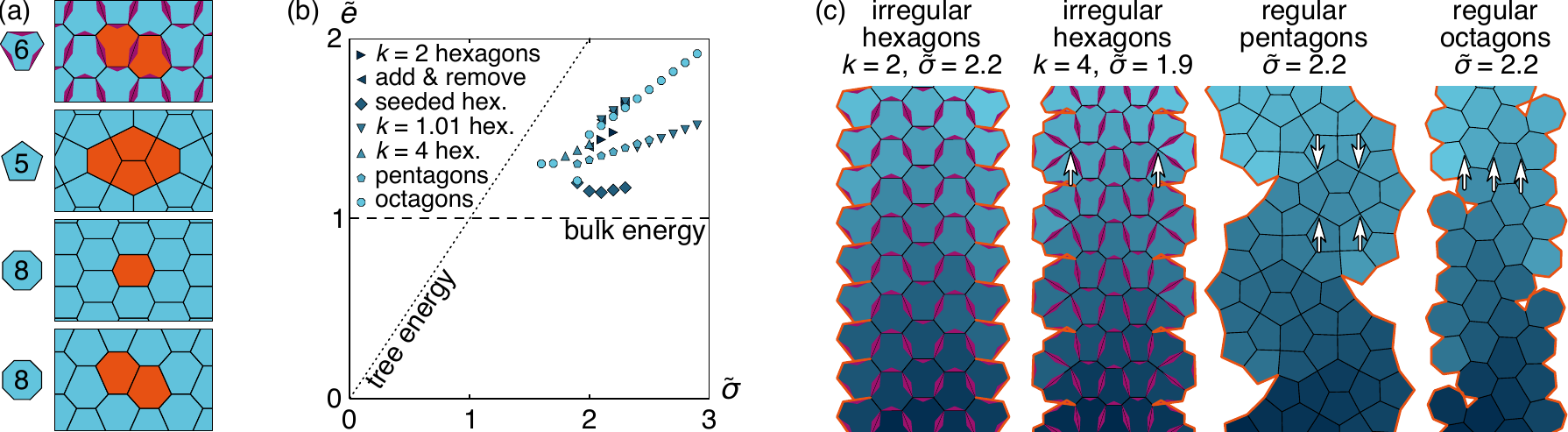}
\caption{\label{fig:closeup}\textbf{Fiber energy and structure.}
(a)~Hexagon, pentagon and octagon bulk structures used to compute the bulk energy $e_b$---and thus the rescaled tension $\tilde{\sigma}$---with unit cells outlined in \emph{orange} (see Methods). Their structures are inspired by aggregates grown under high tensions. The two octagon bulks have identical energies per particle; note that while their polygons superficially appear to have only six sides on the figure, they are actually octagons with two pairs of perfectly aligned sides appearing as longer sides.
(b)~The energy per particle of infinite fibers based on those shown in Fig.~\ref{fig:fibers} is higher that of the bulk, implying that they are nonequilibrium structures.
(c)~Defects, \emph{i.e.}, four-coordinated and two-coordinated vertices, are apparent upon closer inspection of some fibers from Fig.~\ref{fig:fibers} (arrows).
}
\end{figure*}

To test this hypothesis, we simulate irreversible aggregation starting from a single polygon. Our algorithm mimics irreversible protein aggregation, where a particle binding to an existing aggregate does so in the most energetically favorable location without substantial rearrangements of the preexisting aggregate topology. Throughout this process the aggregate energy is always minimized with respect to the position of all its vertices, implying that we impose mechanical force balance within an aggregate before assessing its energy (see Methods).
We first grow aggregates of 150 irregular hexagons. The unfrustrated case $k=1$ simply yields the bulk of Fig.~\ref{fig:model}(a) up to very large values of $\sigma$. Next considering substantially frustrated hexagons with $k=2$, we observe bulks at high ($\tilde{\sigma}\geqslant 2.4$) tensions, while low tensions ($\tilde{\sigma}\leqslant 2.0$) yield irregular tree-like aggregates [Fig.~\ref{fig:fibers}(a)]. By contrast, periodic fibers form at intermediate tensions, and maintain perfect regularity to indefinitely large lengths (see Movie~S1). These fibers form for $\tilde{\sigma}$ of order unity, consistent with our predicted competition between frustration and adhesion. Still, they form closer to $\tilde{\sigma}=2$ than the expected $\tilde{\sigma}=1$, suggesting that fiber formation is not completely captured by equilibrium arguments. To confirm this, we extrapolate the specific energy of our periodic fibers to infinite lengths and compare them to that of the hexagon bulk shown in Figs.~\ref{fig:model}(c) and \ref{fig:closeup}(a). As shown in Fig.~\ref{fig:closeup}(b), the fiber energy exceeds that of the bulk, implying that fibers are indeed out of equilibrium.

To confirm that our fibers are metastable aggregates, we next establish that they are unaffected by small perturbations in the growth pathway but change morphology if nucleated from a more stable phase.
To test the first point, we modify our algorithm to successively add two polygons, then remove one. Similar to polygon addition, our polygon removal procedure minimizes the aggregate energy in a short-sighted fashion, allowing the relaxation of built-up stresses and thus lowering the aggregate energy. The whole procedure is then iterated until an aggregate of the desired size is obtained. As expected, fibers are essentially unaffected by this local change in protocol [Fig.~\ref{fig:fibers}(b)]. We next grow an aggregate from a nucleus of the bulk, inducing significant morphological changes as predicted [Fig.~\ref{fig:fibers}(c)]. The unidirectional, periodic growth is however preserved, attesting to the robustness of the fiber-forming mechanism.

Moving beyond the $k=2$ hexagons considered above, Fig.~\ref{fig:fibers}(d-e) demonstrates that our description is valid for a broad range of $k$ corresponding to variations of the frustration energy $e_b-e_g$ by several orders of magnitude, from $\simeq 7\times 10^{-6}$ for $k=1.01$ to $\simeq 2\times 10^{-1}$ for $k=4$. By comparison the polygon ground state energy $e_g$ remains of order one over this whole range [Fig.~\ref{fig:parametrization}(c)]. Despite these very substantial differences in the magnitude of the frustration energy, the rescaled parameter $\tilde{\sigma}$ remains an excellent predictor of fiber formation. Finally, we move away from hexagons altogether in Fig.~\ref{fig:fibers}(f-g) and demonstrate fiber formation in regular pentagons and octagons, two further shapes that do not tile the plane and thus generate intrinsically frustrated aggregates. Despite very diverse internal fiber structures, the onset of fiber formation is again very well predicted by the criterion $\tilde{\sigma}\simeq 2$.

Our results demonstrate that the inherent geometrical frustration of aggregates of mismatched particles gives rise to a richer range of morphologies than is found in well-matched objects. Most noticeable among these is the robust formation of fibers in regimes where particle adhesion and frustration are comparable in magnitude. Our fibers stand in strong contrast with the three-dimensional morphologies resulting from, \emph{e.g.}, the flocculation of simple spherical colloids. According to our analysis, the formation of slender aggregates is driven by a compromise between, on the one hand, the elastic incentive to place all particles in the vicinity of the boundary of the aggregate to relax their frustrated shapes, and on the other hand the tendency to form a compact aggregate that maximizes adhesion. Though our present demonstration of fiber formation is implemented in two dimensions, the frustration it reveals is also relevant in higher dimensions and will favor the formation of both fibers and sheets in 3D.

The currently dominant paradigm for frustration in soft matter systems assimilates incompatible shapes to a mismatch between an intrinsically curved Riemanian metric favored by the object and the flat metric of the embedding space~\cite{Grason:2016}. These concepts have been successfully applied to (quasi)crystals, surfactant phases or packings of preformed helical fibers, among others~\cite{Sadoc:2008,Bruss:2012,Efrati:2013}. At equilibrium, this mismatch is accomodated by introducing defects in the system~\cite{Bowick:2009} or by forming slender morphologies if defects are strongly penalized~\cite{Schneider:2005,Hure:2011,Meng:2014,Hall:2016,Sharon:2016}. Here slender fibers and topological defects can indeed coexist when using highly frustrated polygons [pentagons, high-$k$ hexagons, octagons; see Fig.~\ref{fig:closeup}(c)]. However our fibers are distinctively out of equilibrium structures, and arise irrespective of whether the intrinsic Gaussian curvature of their constitutive polygons is positive (for pentagons), negative (octagons), or zero (irregular hexagons). These properties contrast with existing Riemanian metric models, suggesting a different frustration mechanism and posing the question of the interplay between the sculpting of the aggregate boundary and the formation of internal defects.

Turning to pathological fiber formation, our results suggest that the distinctive fibrous morphologies of protein aggregates need not be due to a mere coincidental convergence of the underlying molecular mechanisms, but could instead result from generic physical mechanisms. Indeed, while the formation of cross-$\beta$ spines is often discussed as the defining feature of one important class of such fibers, namely amyloids~\cite{Knowles:2014}, deviations from this specific molecular organization have been observed~\cite{Bousset:2002} and secondary interactions have been shown to contribute significantly to their mechanics~\cite{Knowles:2007} and morphologies~\cite{Meinhardt:2009}. These features are consistent with the diverse morphologies obtained in our model upon small variations of our parameters, and could apply to protein fibers with radically different structures~\cite{Eaton:1990}. Beyond biological materials, fiber formation upon aggregation could become a hallmark of self-assembled, frustrated matter, leading to new design principles taking advantage of increasingly sophisticated artificial asymmetrical building blocks at the nano- and micro-scale~\cite{Champion:2007,Wang:2012b}.

\bigskip
\noindent{\normalsize\textbf{\sffamily Acknowledgements}}

\smallskip\small
\noindent We thank Robin Ball for a seminal conversation that inspired this work. We are also grateful to Efi Efrati for his insights on the connection between frustration and intrinsic curvature, and for suggesting our initial fiber-forming realization using pentagons. We thank him and Pierre Ronceray for comments on the manuscript. This work was supported by grants from Universit\'e Paris-Sud's \emph{Attractivit\'e} and CNRS' \emph{PEPS-PTI}  programs, Marie Curie Integration Grant PCIG12-GA-2012-334053, ``Investissements d'Avenir'' LabEx PALM (ANR-10-LABX-0039-PALM), ANR grant ANR-15-CE13-0004-03 and ERC Starting Grant 677532. This work was also supported in part by the National Science Foundation's MRSEC Program under Award Number DMR-1420709. ML's group belongs to the CNRS consortium CellTiss.

\bigskip
\noindent{\normalsize\textbf{\sffamily Methods}}

\smallskip\small
\noindent\textbf{Expression of the aggregate energy.}
The deformation energy for the $\alpha$th polygon is a function of its area $A^{(\alpha)}$ and of the lengths $\lbrace \ell_i^{(\alpha)}\rbrace_{i=1\ldots n}$ of its sides through
\begin{equation}\label{eq:DeformationEnergy}
e^{(\alpha)}=\frac{1}{A^{(\alpha)}}+ \sum_{i=1}^n\frac{k_i}{2}\left[\ell_i^{(\alpha)}\right]^2,
\end{equation}
where $n$ is the number of sides of the polygon.
In this expression, the tendency to extend the area of the polygon due to the first term is counteracted by the harmonic restoring forces due to the second term [Fig.~\ref{fig:parametrization}(b)]. 
The elastic ground state of energy $e_g$ discussed in the text is obtained by minimizing $e^{(\alpha)}$ with respect to the positions of the polygon's vertices.
Ground state polygons are rigid, \emph{i.e.}, devoid of internal soft modes due to the prestress inherent to Eq.~(\ref{eq:DeformationEnergy})~\cite{Alexander:1998}. % [Checked in calculation AES]
The irregularity of this ground state can be continuously tuned through the choice of the spring constants $k_i$, and regular polygons are obtained when they are all identical. The alternating-side polygons presented in Figs.~\ref{fig:model}(b) and \ref{fig:parametrization}(a) and used throughout have $k_1=k_3=k_5=1$ (long sides, marked by \emph{pink tabs}) and $k_2=k_4=k_6=k>1$ (short sides, unmarked), which defines the asymmetry parameter $k$. Throughout this work we concentrate on the domain $k\geq 1$ without loss of generality. Indeed our problem is invariant under the transformation $k\rightarrow 1/k$ by way of a proper rescaling of lengths and energies, and thus the aggregation process at any $k\in(0,1)$ can be inferred from the appropriate $k\in(1,\infty)$. While the specific form of the energy Eq.~(\ref {eq:DeformationEnergy}) is chosen for numerical convenience, its precise expression does not strongly influence our results.

Multiple polygons can be connected through the joining of one or several of their sides. Two joined sides are treated as a single object, implying that they share the same two end-vertices  [Fig.~\ref{fig:parametrization}(a)].
We refer to the specification of all such junctions as the \emph{topology} $\cal T$ of the aggregate. The specification of a $\cal T$ constrains the aggregate shape, and thus tends to increase the deformation energy $e^{(\alpha)}$ of each polygon above $e_g$. In this paper, we consider aggregates whose energies are minimal with respect to the position of their vertices for a given $\cal T$, and denote by $e^{(\alpha)}_{\cal T}$ the deformation energy of particle $\alpha$ in this state of mechanical force balance. We denote by $N_u(\cal T)$ the number of unjoined sides in topology $\cal T$; for instance $N_u=10$ in Fig.~\ref{fig:parametrization}(a), as indicated by the orange lines. Imposing the surface tension energy penalty $\sigma>0$ to each unjoined side regardless of its actual length, the total energy of an aggregate comprising $N$ polygons thus reads
\begin{equation}\label{eq:AggregateEnergy}
E(\lbrace k_i\rbrace, \sigma,{\cal T}) = \sum_{\alpha=1}^N \left[e^{(\alpha)}_{\cal T}-e_g\right]
+ N_u ({\cal T})\sigma.
\end{equation}
The first term of the right-hand side of Eq.~(\ref{eq:AggregateEnergy}) describes the total deformation energy in excess of the ground state energy $Ne_g$, while the second term is the surface energy. 
Overall, $E$ depends on the structure of the aggregate only through its topology $\cal T$. The average energy per particle discussed in the text is defined as $e=E/N$.

\medskip

\noindent\textbf{Surface energy for trees and bulks.} We compute the surface energy of a tree comprising $N$ polygons by noting that it has $N_u=2+N(n-2)$ unjoined sides. We demonstrate this by recursion over $N$, noting that $N_u=n$ for $N=1$. Each additional polygon adds $n$ sides to the existing aggregate, $n-1$ of which are unjoined. The existing aggregate moreover loses one unjoined side to the connection with the new polygon. Thus $N_u$ is incremented by $n-2$ each time a new polygon is added to the tree, which proves our statement. The number of unjoined sides in a tree is thus proportional to its total number of polygons. As the number of unjoined sides in a two-dimensional bulk is proportional to $\sqrt{N}$, in the thermodynamic limit its surface energy is negligible compared to its finite elastic energy per particle.

\medskip

\begin{figure}[t]
\includegraphics[width=\columnwidth]{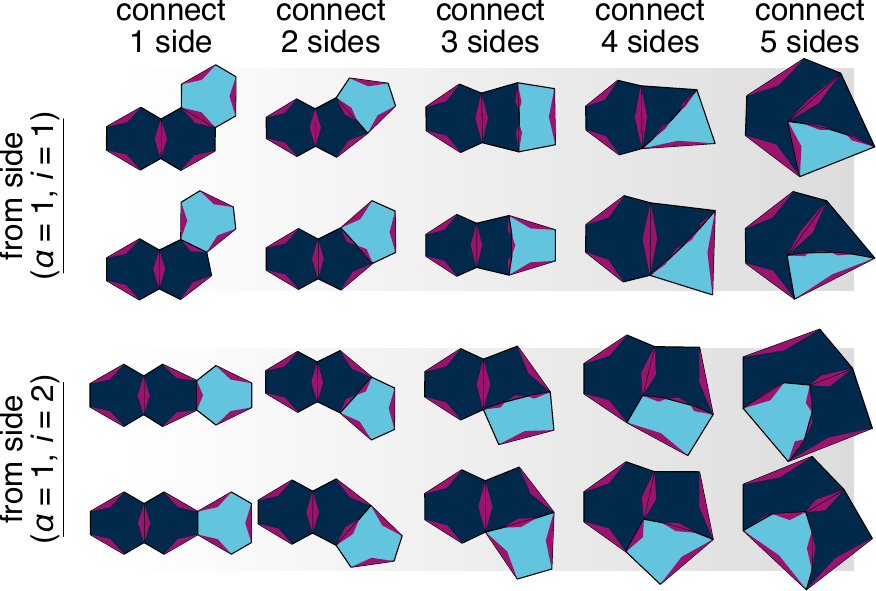}
\caption{\label{fig:algorithm}\textbf{Aggregation algorithm.} Our aggregate growth algorithm enumerates the complete set of topologies obtained by adding a new polygon (\emph{light blue}) to an existing aggregate (\emph{dark blue}) identical to that of Fig.~\ref{fig:parametrization}(a). Here we illustrate only the 20 topologies obtained by adding a new polygon to sides $(\alpha=1,i=1)$ and $(\alpha=1,i=2)$ of the existing aggregate. Our algorithm considers 80 more topologies associated with the remaining eight sides.}
\end{figure}

\noindent\textbf{Deformation energy for trees and bulks.} The deformation energy of trees and bulks are computed by minimizing the energy Eq.~(\ref{eq:DeformationEnergy}) respectively with free boundary conditions or assuming the lattice structures illustrated in Fig.~\ref{fig:closeup}(a) for regular pentagons:
\begin{equation}
e_g=2\sqrt{2}\left(5-2\sqrt{5}\right)^{1/4}, \quad e_b=\sqrt{6},
\end{equation}
regular octagons:
\begin{equation}
e_g=2\sqrt{2(\sqrt{2}-1)},\quad e_b=\frac{2^{3/2}}{5^{1/4}},
\end{equation}
and irregular hexagons:
\begin{eqnarray}
e_g&=&\frac{2\sqrt{2}}{3 b+\sqrt{3} a}\left\lbrace\sqrt{3} (k-2) a^2+2 \sqrt{3} (k+1) a\right.\nonumber\\
&&\left. -3 b \left[2 a+\sqrt{3} k b-2 (k+1)\right]\right\rbrace^{1/2},\nonumber\\
e_b&=&\frac{2^{3/2}k^{1/4}}{3^{1/4}},
\end{eqnarray}
with $a=\left(3-3k+\sqrt{1+14k+k^2}\right)/\left(4+4k\right)$ and $b=\sqrt{1-a^2}$.

\medskip

\noindent\textbf{Sequential aggregation algorithm.}
We use a deterministic algorithm that considers all possible options for the addition of a polygon onto an existing aggregate, some of which are illustrated in Fig.~\ref{fig:algorithm}. After minimizing the total energy of the whole aggregate with respect to the coordinates of all its vertices for each option (which induces the polygon distortions seen in the figure), it selects the option associated with the lowest total energy $E$ and uses the result as the basis of the next polygon addition.
Similar to kinetic, irreversible protein aggregation \emph{in vivo}, this procedure thus does not necessarily achieve the most energetically favorable aggregate topology globally. Instead, our algorithm locally guarantees the best energetic choice at each addition step, allowing the formation of metastable aggregates characteristic of our frustrated interactions.

\medskip

\noindent\textbf{Movie S1.}
Characteristic growth sequence of a large fiber suggesting that periodic fibers elongate indefinitely. Here $N=450$ for $k=2$ irregular hexagon four-periodic fiber with $\tilde{\sigma}=2.2$. The scale bar in the left-hand side has unit length in simulation units. %1401_0021}

\end{document}